Research and Applications

# Using Twitter data to understand public perceptions of approved versus off-label use for COVID-19-related medications

Yining Hua[1,2], Hang Jiang[3], Shixu Lin[4], Jie Yang[4], Joseph M. Plasek[1,2], David W. Bates [1,2], and Li Zhou[1,2]

[1]Department of Biomedical Informatics, Harvard Medical School, Boston, Massachusetts, USA, [2]Division of General Internal Medicine and Primary Care, Brigham and Women's Hospital, Boston, Massachusetts, USA, [3]Massachusetts Institute of Technology, Cambridge, Massachusetts, USA, and [4]Zhejiang University School of Medicine, Hangzhou, Zhejiang, China

Yining Hua and Hang Jiang contributed equally to this work.

Corresponding Author: Yining Hua, BA, Division of General Internal Medicine and Primary Care, Brigham and Women's Hospital, 399 Revolution Dr, Suite 1315, Somerville, MA 02145, USA; yining_hua@hms.harvard.edu

## ABSTRACT

**Objective:** Understanding public discourse on emergency use of unproven therapeutics is essential to monitor safe use and combat misinformation. We developed a natural language processing-based pipeline to understand public perceptions of and stances on coronavirus disease 2019 (COVID-19)-related drugs on Twitter across time.

**Methods:** This retrospective study included 609 189 US-based tweets between January 29, 2020 and November 30, 2021 on 4 drugs that gained wide public attention during the COVID-19 pandemic: (1) Hydroxychloroquine and Ivermectin, drug therapies with anecdotal evidence; and (2) Molnupiravir and Remdesivir, FDA-approved treatment options for eligible patients. Time-trend analysis was used to understand the popularity and related events. Content and demographic analyses were conducted to explore potential rationales of people's stances on each drug.

**Results:** Time-trend analysis revealed that Hydroxychloroquine and Ivermectin received much more discussion than Molnupiravir and Remdesivir, particularly during COVID-19 surges. Hydroxychloroquine and Ivermectin were highly politicized, related to conspiracy theories, hearsay, celebrity effects, etc. The distribution of stance between the 2 major US political parties was significantly different ($P < .001$); Republicans were much more likely to support Hydroxychloroquine (+55%) and Ivermectin (+30%) than Democrats. People with healthcare backgrounds tended to oppose Hydroxychloroquine (+7%) more than the general population; in contrast, the general population was more likely to support Ivermectin (+14%).

**Conclusion:** Our study found that social media users with have different perceptions and stances on off-label versus FDA-authorized drug use across different stages of COVID-19, indicating that health systems, regulatory agencies, and policymakers should design tailored strategies to monitor and reduce misinformation for promoting safe drug use. Our analysis pipeline and stance detection models are made public at https://github.com/ningkko/COVID-drug.

## INTRODUCTION

The emergence of the novel pathogen coronavirus disease 2019 (COVID-19) (SARS-CoV-2) resulted in the rapid publication of clinical data. As expected, these data produced preliminary and often anecdotal or underpowered evidence in the early stages of the pandemic. These preliminary data were widely used to promote the off-label use of therapeutics within social media platforms (such as Twitter) before US Food and Drug Administration (FDA) Emergency Use Authorizations (EUAs)[1,2] or sufficient safety and efficacy evidence such as randomized controlled trials (RCTs). Off-label drug use, or the unproven use of an approved drug, has been used on a large number of patients as a "last resort" before COVID-19-targeted therapies were developed. However, off-label drug use can be potentially dangerous because of adverse effects. For example, chloroquine and hydroxychloroquine, which were largely prescribed in the early-pandemic period, can cause serious heart rhythm problems (eg, QT prolongation, tachycardia), leading to an increased risk of cardiac death.[3,4] The harm from unproven, off-label use may also outweigh the benefits when considering that the most vulnerable groups for COVID-19 are the elderly or patients with multiple comorbidities.[5,6] Given the fact that widespread off-label and compassionate use of drugs for COVID-19 can harm the public and also discourage patients and clinicians from participating in RCTs,[6] understanding the public's attitudes toward off-label drug use and why people may promote off-label drug use during such times of anxiety is critical to promoting safe medication use.

Studies have used social media data to understand public opinions on COVID-19-related topics and detect misinformation disseminated via social media platforms.[7] Despite a growing body of studies that have used Twitter data to study COVID-19-related issues and events, studies on drug use have so far only focused on Hydroxychloroquine, Remdesivir, and Convalescent plasma therapy, with a major focus on Hydroxychloroquine.[8–10] However, most research to date has not leveraged modern natural language processing (NLP) techniques (eg, BERT[11]) to extract fine-grained opinions or investigate the relationship between stance and multiple social factors. To fill in gaps in this research area, we aimed to understand and compare public perception of 4 different drugs on Twitter during different waves of the pandemic, specifically: (1) Hydroxychloroquine and Ivermectin, drugs with anecdotal evidence in preventing and treating COVID-19; and (2) Molnupiravir and Remdesivir, FDA-approved treatment options for eligible patients with COVID-19. We developed an analytic pipeline that integrates advanced computational technologies in NLP and machine learning to investigate the public's standpoint on the 4 drugs, and its relationship with age, medical background, and political leanings over time using 2 years of Twitter data. Specifically, we aimed to answer the following research questions:

1. Were off-label drugs more popularly discussed than COVID-19-targeted drugs under development?
2. What did people discuss about these drugs? What were their supporting or opposing rationales?
3. Who was prone to support emergency use of unauthorized off-label drugs versus FDA-authorized drugs?

The first question can be answered by tracking the number of tweets related to each drug over time. Similar methods have been used to track the popularity of different study subjects.[12–15] The second question usually requires NLP techniques to detect stance and conduct a content analysis on tweets. Stance detection is an approach to automatically detect an author's stance (eg, support, against, neutral) towards a specific topic. It has been applied to study the public perception of COVID-19 vaccines,[16–18] mask-wearing,[15,19] and other COVID-19 topics. This study adopted a variant of BERT[11] to embed documents for clustering and topic generation, which has been tested and used in COVID-19 research.[20–23] Last, we applied geospatial and demographic analysis to answer the third question. Previous work has established the validity of using demographic analysis to understand disparities among various population groups (eg, gender, age) in concern and their sentiments towards vaccines or mask-wearing during the pandemic.[25–27]

## MATERIALS AND METHODS

### Study design

In this retrospective cohort study, we developed a pipeline using NLP, machine learning, and statistical methods to conduct the following analyses using Twitter data (Figure 1): (1) a time-trend analysis to examine the relationship between the drug-related tweets and the number of newly confirmed COVID-19 cases and public stance on each drug; (2) demographic analyses to understand the public sentiment on the drugs and demographic attributes of different groups of people; and (3) viewpoint extraction with content analyses to study detailed rationales behind people's perceptions on the drugs.

This study has been approved by the Mass General Brigham International Review Board.

### Data sources

Two publicly available datasets, John Hopkins University's (JHU) COVID-19 data repository (JHU-CDR)[25] and the COVID19_Tweets_Dataset (CTD),[26] were used in this study. JHU-CDR contains global confirmed COVID-19 daily case counts for each region or country. CTD contains unique tweet identifiers (IDs) associated with COVID-19 and is mapped to the tweet IDs in previously published research on this topic by Chen et al.[27] Figure 2 shows the number of tweets in each step of the study.

### Data preparation

This study included only original (nonreposted) English tweets originating in the United States. Ninety-three weeks of tweets from January 29, 2020, 1 week after the first laboratory-confirmed COVID-19 case report in the United States,[24] to November 30, 2021. A total of 396 380 339 English COVID-19-related tweets were pulled. The 93 weeks were divided into 3 sections based on the onset of the 3 major waves of COVID-19 in the United States, which were determined based on the weekly distribution of new cases. Wave 1 starts on January 29, 2020, the first date in the dataset. We set the beginnings for wave 2 (September 16, 2020) and wave 3 (July 7, 2021) to be the first rebound of new cases after a constant declining pattern (Figure 3).

We focused on 4 drugs that received relatively high public attention during the pandemic (ie, Hydroxychloroquine, Ivermectin, Molnupiravir, and Remdesivir; Table 1). A drug lexicon (Supplementary Appendix SA) was curated by clinical informaticians to case-insensitively extract tweets mentioning the 4 drugs. The drug lexicon was also extended via manual review of candidates that were fuzzy matched to appropriate synonyms. The drug-related tweets were further extracted to include only tweets with geolocation in the United States (Supplementary Appendix SB). In total, we identified 108 976 Twitter users with 609 189 tweets.

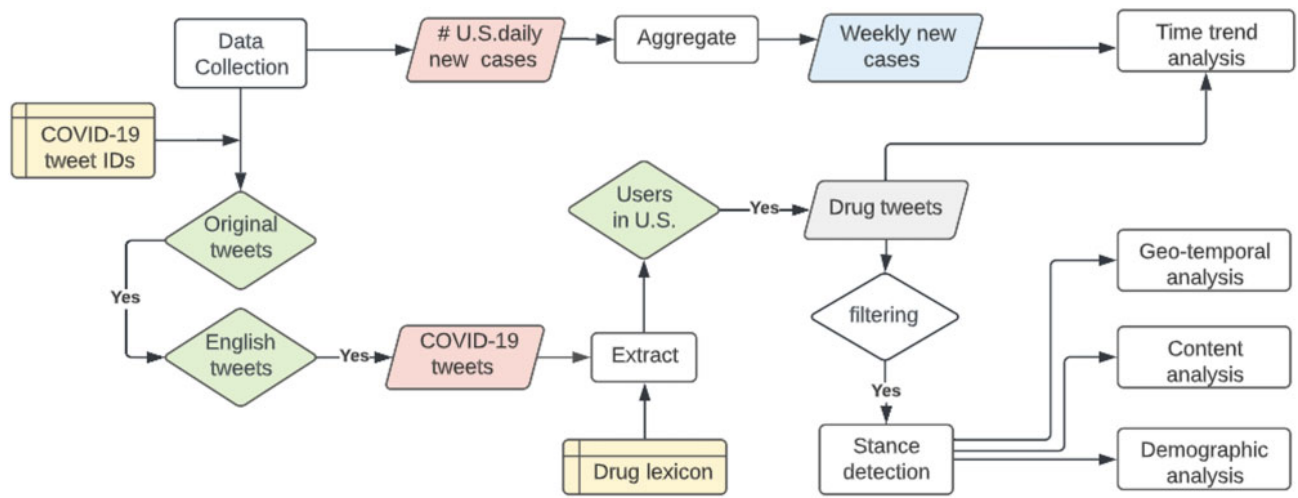

**Figure 1.** A comprehensive multimodal pipeline to study the public perception of drugs during the COVID-19 period.

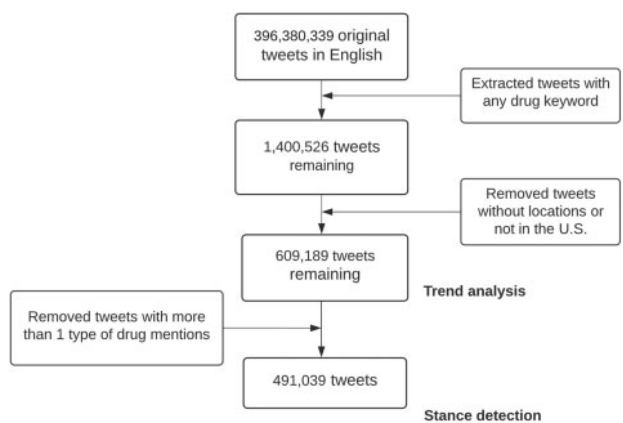

**Figure 2.** Data flow diagram.

### Time trend analysis and correlation study
Daily new COVID-19 case counts were aggregated into the same weekly metric as the tweets (Tuesday–Monday). The weekly number of new cases and numbers of tweets related to each drug is overlaid to compare trends in public attention on each drug across time.

### Stance detection model
We developed 2 COVID-drug-stance RoBERTa-base models by fine-tuning a pretrained Twitter-specific stance detection model[30] on a stance dataset called COVID-CQ.[31] COVID-CQ contains 3-label annotated opinions (negative, neutral, and positive; with 14 353, 14 373, and 14 372 tweets, respectively) of the tweet initiators regarding the use of Chloroquine or Hydroxychloroquine for the treatment or prevention of the coronavirus. The data were divided into training and validation datasets with a 70:30 ratio. Model I (COVID-drug-stance-BERT) was trained on the original tweet data, and Model II (COVID-drug-stance-BERT-masked) was trained on tweets with drug names masked as "[mask]" for model generalizability on different drugs. The 2 models had similar performance on the COVID-19 validation set: COVID-drug-stance-BERT had an accuracy of 86.88%, and the masked model had an accuracy of 86.67%. The 2 models were then evaluated by predicting tweet initiators' attitudes towards the drug mentioned in each tweet using randomly selected test sets (100 tweets) of each drug. Tweets with more than 1 drug mention were removed in this step and for later stance-based analysis (number of tweets summarized in Supplementary Appendix SC). As suggested by the evaluation in Table 2, Model I had better performance and was therefore used in this study.

### Content analysis
Here, we propose a new pipeline for content analysis. We used text clustering and Named Entity Recognition (NER) word clouds to summarize public opinions for content analysis. For text-clustering, we used BERTweet to embed tweets and principal component analysis to reduce the embedding dimensions. We classified tweets based on drugs and stances and used K-Means to find $k$ clusters for each drug per wave. Then, we used SentenceBERT to find $n$ closest sentences in the same clusters and summarized the main arguments for supporting using the drugs as cures or preventatives for COVID-19. We used the elbow method and empirically set $k = 15$ and $n = 30$ via a manual review process (Supplementary Appendix SD).

For NER word clouds, Stanza's 4-class (person, organization, location, and miscellaneous) tweet NER model was used to extract named entities.[32,33] The extracted NERs were plotted per wave per drug to show the change in people's foci over time.

### Demographic analysis
We conducted demographic inference, including geolocation, political partisanship, healthcare background, age, and gender in this study (Supplementary Appendix SE). In short, we calculated and visualized statewide average stance per drug per wave for geoinference; We follow Demszky et al[34] to infer political affiliation and Li et al[35] to infer healthcare background. To infer age and gender, we used a standard approach called M3 as presented by Wang et al.[36] We calculated and visualized the distribution of each group as classified by their stances. Finally, we used the Pearson Chi-Square test to examine if there were significant associations between demographic groups and different attitudes towards the drugs.

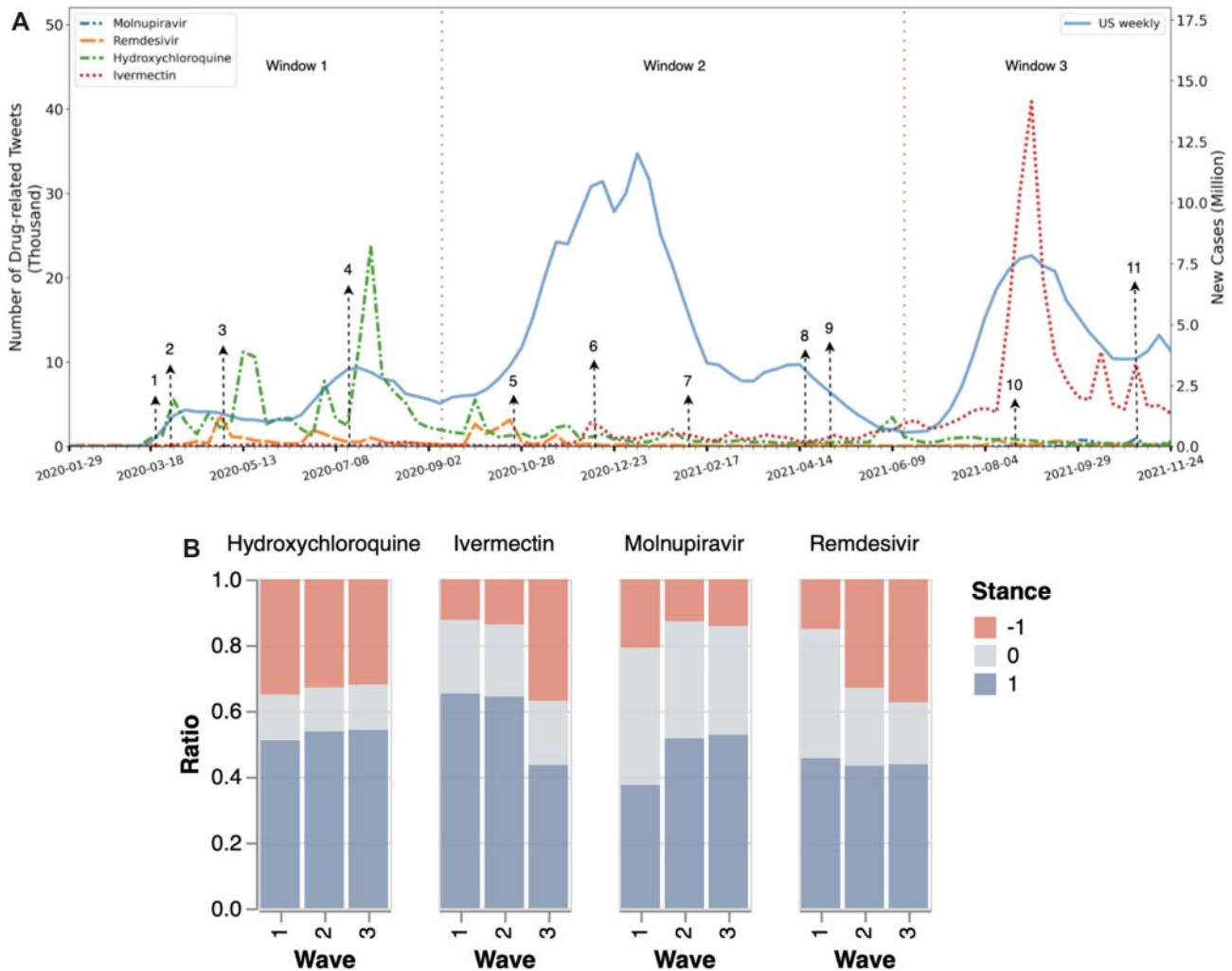

**Figure 3.** (A) The trends of (1) the number of tweets that mentioned COVID-19-related drugs: Hydroxychloroquine, Ivermectin, Molnupiravir, Remdesivir, and (2) weekly COVID-19 case counts (stepped line) in the United States. Wave boundaries are noted by dashed vertical lines. Major drug events are noted by numbers: (1) March 19, 2020: Trump declared Hydroxychloroquine a game-changer; (2) March 28, 2020: FDA approved a EUA to use Hydroxychloroquine for certain hospitalized patients; (3) May 1, 2020: FDA approved a EUA to use Remdesivir for severe patients; (4) July 15, 2020: FDA cautioned against the use of Hydroxychloroquine; (5) October 22, 2020: FDA approved Remdesivir for conditional use; (6) December 10, 2020: FDA cautioned against Ivermectin; (7) February 4, 2021: Merck cautioned against Ivermectin; (8) April 17, 2021: FDA clarified that Remdesivir was not approved; (9) May 1, 2021: FDA recalled a batch of Remdesivir vials, (10) August 21, 2021: FDA denounced Ivermectin as a COVID-19 treatment following an increase in overdoses; (11) November 4, 2021 Britain authorized Molnupiravir for COVID-19 treatment. (B) Distribution of percentage of tweets with positive (1, blue), neutral (0, gray), and negative (−1, red) stances for each drug. COVID-19: coronavirus disease 2019; EUA: Emergency Use Authorization FDA: US Food and Drug Administration.

## RESULTS

### Were off-label drugs more popularly discussed than COVID-19 drugs underdevelopment?

Among 609 189 tweets geotagged in the US, the public interest was primarily in Hydroxychloroquine and Ivermectin (255 573, 42.0% and 332 381, 54.6% mentions, respectively), and most tweets showed a supportive attitude (53.04% and 57.72%, respectively) across all 3 waves (Figure 3). Hydroxychloroquine received the most attention in the first wave and had an increasing proportion of supporters and a decreasing number of opposers across the 3 waves. In contrast, Ivermectin gained popularity in the third wave but had a very high approval and low disapproval rate in the first 2 waves. In wave 3, Ivermectin had a much lower approval rate and a more than 2-fold increase in the disapproval rate. We hypothesize that the increase in disapproval may have been a consequence of the FDA denouncing Ivermectin as a COVID-19 treatment due to increasing overdose rate (event 10 in Figure 3). In contrast, Molnupiravir and Remdesivir received sporadic attention during the study period (6285, 1.0% and 54 950, 9.0% mentions, respectively), and had a lower proportion of supporters (47.32%, 44.27%). Molnupiravir only started to receive some attention after September 2021, and it had the highest ratio of neutral users, though its supporters had been gradually increasing. Public discussion concerning Remdesivir on Twitter dramatically decreased after its conditional FDA approval on October 22, 2020, at the beginning of wave 2 (event 5). It continued to have an increasing number of opponents, but the proportion of supporters was almost flat.

### What did people discuss about the drugs?

Through NER (Figure 4), we found that Hydroxychloroquine had a clear and consistent connection with the former US President Donald Trump, and his endorsement of the therapy.[37] Ivermectin, in contrast, was linked to different named entities throughout 3 waves, such as people (eg, Thomas Borody, Pierre Kory, and Joe Rogan) who proposed or advocated the use of Ivermectin for COVID-19 and places (eg, Australia, India) where it was used in a controversial manner. Molnupiravir, co-developed by Merck and Ridgeback, had been associated with these 2 organizations in all 3 waves (eg, Julie Gerberding, the Chief Patient Officer and Executive Vice President at Merck). Besides, Pfizer's oral antiviral named Paxlovid which showed superior effectiveness in clinical trials was also mentioned in both waves 2 and 3. The UK and Britain were also mentioned in relation to Molnupiravir in wave 3 because it was approved there. Remdesivir, invented by Gilead Science, was likewise linked to the company in all 3 waves. Additionally, Remdesivir was also consistently associated with famous people who supported use of the drug (eg, Anthony Fauci, Donald Trump, and Rand Paul).

**Table 1.** Treatment indications[28] and FDA approval information with dates[29] of the selected drugs

| Drug name | Indications and usage | FDA approval information and dates (regarding COVID-19) |
| --- | --- | --- |
| **Hydroxychloroquine** | Prevent and treat acute attacks of malaria. Treat rheumatoid arthritis and systemic lupus erythematosus. | March 28, 2020: FDA issued an Emergency Use Authorization (EUA). June 15, 2020: EUA canceled. |
| **Ivermectin** | Treat intestinal parasitic worms, parasitic diseases, Strongyloides Stercoralis and Onchocerca Volvulus. | Never been approved. |
| **Molnupiravir** | Treat both influenza and COVID-19 for patients infected by SARS-CoV-2. | December 23, 2021: EUA issued. March 20, 2020: Compassionate use approved. May 1, 2020: EUA issued. October 22, 2020: Approved for adults and pediatric patients 12 years of age and older. |

**Table 2.** Evaluation of 2 drug-stance models on each of the 4 drugs

| | Model I: original tweet | | | Model II: drug names masked | | |
| --- | --- | --- | --- | --- | --- | --- |
| Drug | Precision | Recall | F1-Score | Precision | Recall | F1-Score |
| Hydroxychloroquine | 0.93 | 0.92 | 0.92 | 0.84 | 0.83 | 0.83 |
| Ivermectin | 0.92 | 0.91 | 0.91 | 0.72 | 0.68 | 0.68 |
| Molnupiravir | 0.89 | 0.89 | 0.89 | 0.78 | 0.77 | 0.77 |
| Remdesivir | 0.82 | 0.79 | 0.79 | 0.70 | 0.66 | 0.66 |

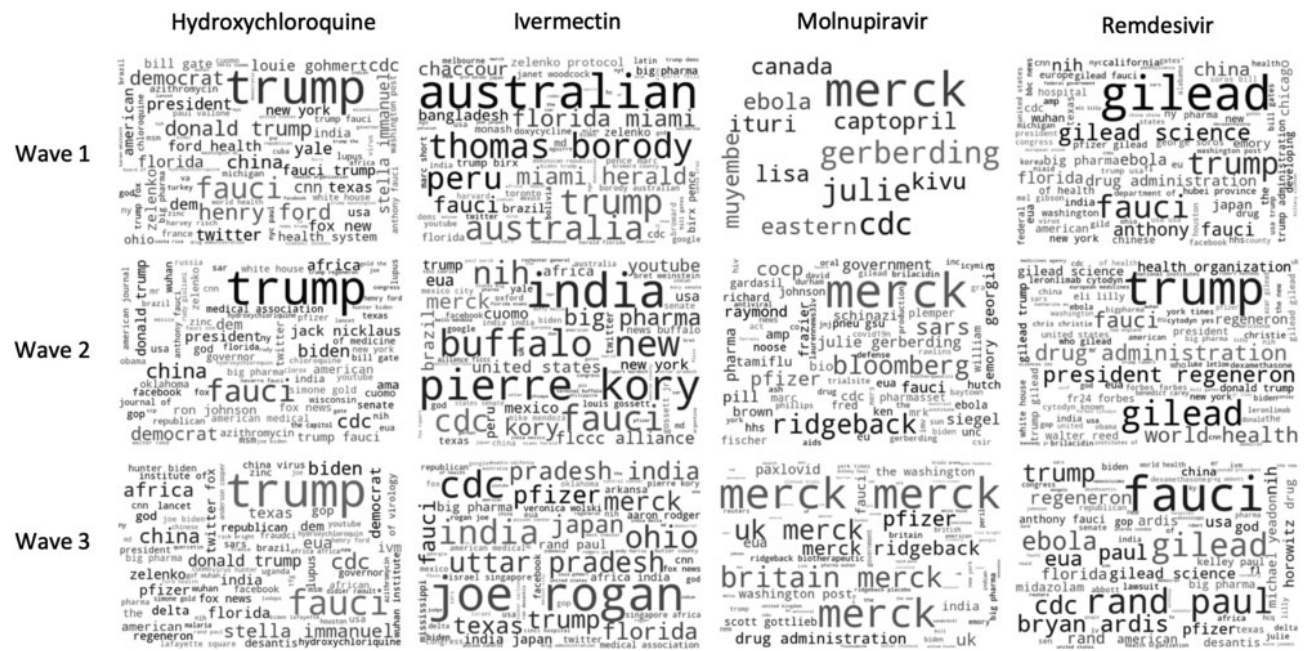

**Figure 4.** Word clouds of named entities per drug per wave. The stances and interests of influencers such as politicians, doctors, and celebrities as well as the relevant pharmaceutical companies that developed the drugs were usually the centers of the discussions.

**Table 3.** Top 5 user rationales for supporting each drug[a]

| Top rationales | | | |
| --- | --- | --- | --- |
| Hydroxychloroquine | Ivermectin | Molnupiravir | Remdesivir |
| Supported by scientific research, health workers, and doctors | Supported by scientific research, health workers and doctors | Supported by scientific research, health workers and doctors | Supported by scientific research, health workers, and doctors |
| Effective shown by experiences of celebrity/family/friends/self/people in other countries | Effective shown by experiences of celebrity/family/friends/self/people in other countries | Produced at a low cost | Effective shown by experiences of celebrity/family/friends/self/people in other countries |
| Cheap, banned because pharmaceutical companies and Dr. Anthony Fauci are lucrative. | Cheap, banned because pharmaceutical companies and Dr. Anthony Fauci are lucrative. | Approved by the UK regulatory authorities | Issued an Emergency Use Approval by the UK regulatory authorities |
| Safely used for decades, safer than vaccines | Safely used for decades, safer than vaccines | Recommend by FDA advisers for authorization | Issued an Emergency Use Approval by the FDA |
| Deliberately politicized by Democrats to oppose Trump | Never tested by institutions on humans because they know it will work | Tested by Georgia State University and recommended for Emergency Use Approval | Supported by both the Trump administration and Dr. Anthony Fauci |

[a]The same rationales across different drugs are highlighted using the same color.

The content analysis results summarized in Table 3 show the top 5 arguments from supporters of each drug. The top rationales for supporting the off-label drugs concerned authoritative evidence, anecdotal evidence, price, safety, and politics, while the top rationales for supporting the COVID-19 targeted drugs exclusively concerned authoritative evidence, anecdotal evidence, and policies. Hydroxychloroquine and Ivermectin supporters had more rationales than supporters of Molnupiravir and Remdesivir.

For Hydroxychloroquine and Ivermectin, a large proportion of the proponents cited certain evidence from published research studies, health workers, and doctors. However, most people cite the sources regardless of their factuality. For example, some people cited clinical results without RCTs or given by research that was later retracted. A group of proponents for Hydroxychloroquine and Ivermectin argued that India and numerous other countries such as Australia and Japan approved these 2 drugs to treat COVID-19 and related the mitigating trends in these countries to these drugs. However, India has de-approved the use of the 2 drugs since September 24, 2021,[38] while Australia and Japan have never authorized Ivermectin to treat COVID-19.[39–41] Other users believed that Hydroxychloroquine and Ivermectin were safer than vaccines because they had been safely used for a long time, while vaccines were recently developed and therefore unreliable. In accordance with previous literature, a group of users liked to cite anecdotal evidence from celebrities, people around them, or themselves. We also observed that Hydroxychloroquine was highly politicized, whereas both Molnupiravir and Remdesivir proponents had few statements. They tended to draw evidence from recent news, reports, UK regulations, or FDA regulations to support their views. Interestingly, some people supported Molnupiravir because they believed that Merck is a good company for not trying to profit from their drug, Ivermectin. This contrasts sharply with Remdesivir because the main argument from its opponents was that the authorities promoted Remdesivir over cheaper therapeutics in order to make a profit.

We also analyzed the reasons given by the opponents of each drug but found that there were only a few reasons, and they were mostly retaliatory statements to the reasons given by the proponents. Some examples of the reasons are that experiments showed that the drug was ineffective, expensive, and incorrectly considered effective for political reasons.

### Who was prone to support emergency use of unauthorized off-label drugs versus FDA-authorized drugs?

Figure 5 demonstrates that the stance towards each drug varied state by state and changed across COVID waves. The overall attitude towards Hydroxychloroquine and Ivermectin changed from positive to neutral, opposite to that of Molnupiravir and Remdesivir. Hydroxychloroquine had an average positive sentiment manifested in Wyoming and Alabama in waves 1–2, and Louisiana in wave 3. Ivermectin had positive attitudes primarily in Montana, Arizona, Florida, etc., but only in waves 1–2. Molnupiravir had consistent positive discussions in most states that had data, while Remdesivir had mostly neutral discussions except in wave 3.

Figure 6 shows the results of demographic analyses. The political partisanship assignment method predicts 45.5% (Democratic), 16.3% (Neutral), and 38.3% (Republican) from the total sample. The distribution of stance for left and right partisans was significantly different (Chi-square test) for the 4 drugs ($P < .001$ for all). The result also suggested that the Republicans were more likely to support Hydroxychloroquine (71% vs. 16%) and Ivermectin (63% vs. 33%) than Democrats. This corresponds to our findings in the content analysis: Hydroxychloroquine supporters thought that the therapy was politicized by the Democrats as a measure to oppose the former president. It is also expected that attitude regarding Remdesivir was polarized as our content analysis shows that some Hydroxychloroquine and Ivermectin supporters believed that Anthony Fauci was involved with the invention of and profited from Remdesivir.

For healthcare backgrounds, we found that around 9.0% tweets came from users with healthcare backgrounds. The stance distribution of users with healthcare background significantly differed from the other group for all drugs ($P < .001$) but Molnupiravir ($P = .029$). The analysis also shows that users with healthcare backgrounds were more likely to support Molnupiravir (49.3% vs. 42.0%) and oppose Hydroxychloroquine (48.0% vs. 41.7%) compared to the general population, while the general population tended to support Ivermectin (49.5% vs. 36.0%) and Remdesivir (44.5% vs. 41.2%).

Regarding gender (Supplementary Appendix SF), 71.8% of tweets came from male users and 28.2% came from female users, which was different from the standard female (43.6%) and male (56.4%) ratio on Twitter.[42] The stance distribution for predicted gender is significantly different between male and female users for Hydroxychloroquine, Ivermectin, and Remdesivir ($P < .001$), but not Molnupiravir ($P = .036$).

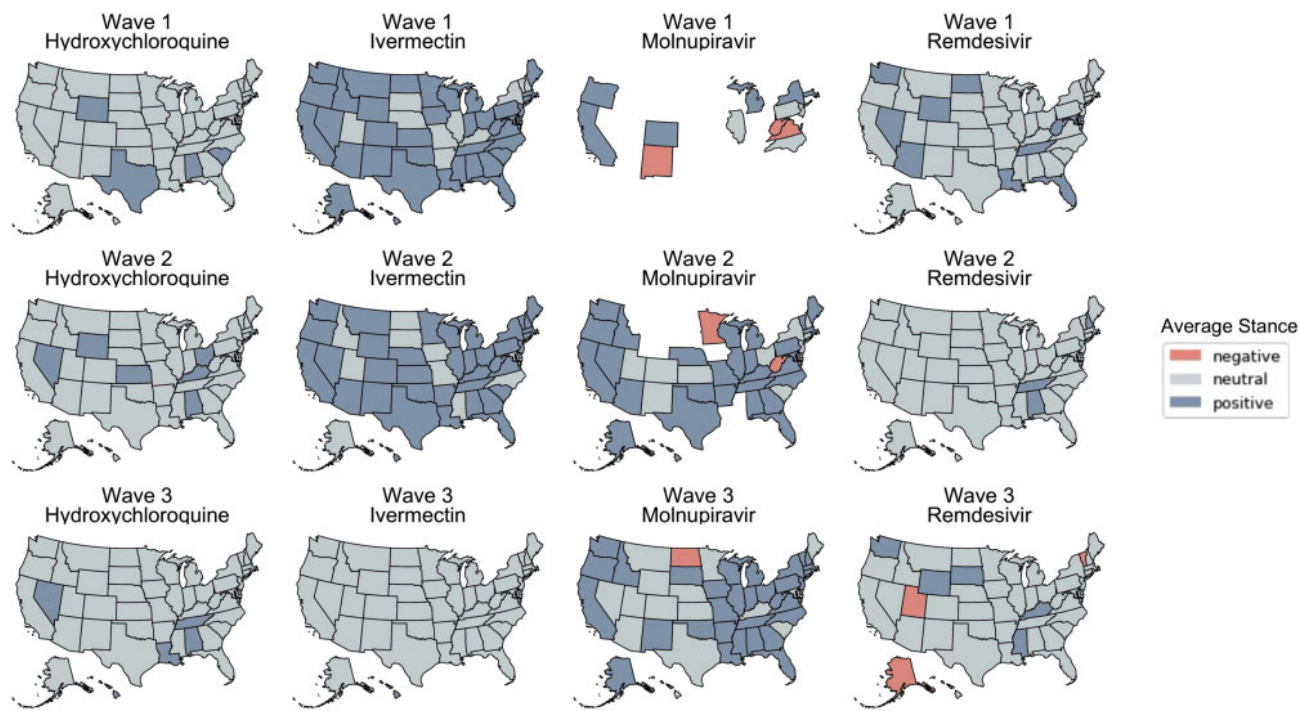

**Figure 5.** Longitudinal geo-temporal analysis of Tweeted sentiment of the 4 drugs by COVID-19 pandemic wave. The average sentiment of each state was classified into positive, neutral, and negative. COVID-19: coronavirus disease 2019.

Finally, the age distribution for all the tweets is 7.1% ($\leq$18), 28.5% (19–39), and 64.5% ($\geq$40). Considering that Twitter users mainly consist of younger users (18–29) as suggested by a census in 2021,[43] we can infer that the age group of $\geq$40 had been the main force in the drug-related discussion. The age plot showed that there is no obvious relationship between age and stance for the 4 drugs (Supplementary Appendix SF).

## DISCUSSION

In this study, we leveraged a large amount of Twitter data (396 380 339 original tweets across 93 weeks), advanced NLP (BERT), and machine learning technologies (K-Means-based topic clustering) to assess public perception of COVID-19-related drugs across time. We developed and released an NLP pipeline based on BERT stance detection and text clustering models for content and demographic analyses. To the best of our knowledge, this is the first work to (1) develop and release stance detection models on social media discourse on drugs; (2) apply stance detection models to characterize group attitudes toward COVID-19 drugs; (3) study the relationship between public stances and the underlying demographic features, regarding 4 popular COVID-19-related drugs. Through this pipeline, we successfully (1) compared the attention received by 4 COVID-19-related drugs: 2 repurposed off-label drugs (Hydroxychloroquine and Ivermectin) and 2 drugs authorized by FDA for COVID-19 (Molnupiravir and Remdesivir), across 3 pandemic waves; (2) investigated the potential impacts of milestone events on the public discussions on drug use; and (3) compared demographic factors (geolocation, political orientation, medical background, age, and gender) that potentially affected tweet frequency and stance distribution of each drug.

We found that although off-label drug use is risky, the public still repeatedly supported taking the risk when there were no present treatments for disease. Repurposed drugs which only have anecdotal evidence were more popularly discussed and supported compared to drugs that were developed for COVID-19 or have been approved by the FDA for emergency use without retraction. Also, through the NER analysis, we found that celebrities and political figures such as the comedian Joe Rogan and former President Donald Trump appeared far more influential than others. For example, Anthony Fauci, the Chief Medical Advisor on COVID-19 to the President, was frequently mentioned suggesting that he suppressed Hydroxychloroquine for his own benefit and pushed the expensive new drug Remdesivir, in which he has a financial stake. This is in accordance with a previous study that shows that "fake news" and inaccurate information may spread faster and more widely than fact-based news. The celebrity effect and conspiracy theories allow misinformation to disseminate more broadly and be believed without significant evidence.[44,45] For example, just 2 days after the March 19, 2020 press conference held by Donald Trump, purchases for medicine substitutes such as Hydroxychloroquine had increased by 200%.[37] Through content analysis, we also found factors that contributed to the popularity of Hydroxychloroquine and Ivermectin: (1) the affordable price, (2) overconfidence in its safety (eg, believing that the worst outcome is only ineffectiveness), (3) hearing about other people's good experiences with it and believing that other countries have seen improved COVID-19 outcomes because they used the drug, and even (4) the belief that the authorities are keeping the drug out of the public's hands for their own benefit.

We noticed that some groups of people believed that Democrats used Hydroxychloroquine to defame Donald Trump and the Republican party. Some users believed that the Democrats tried to ban

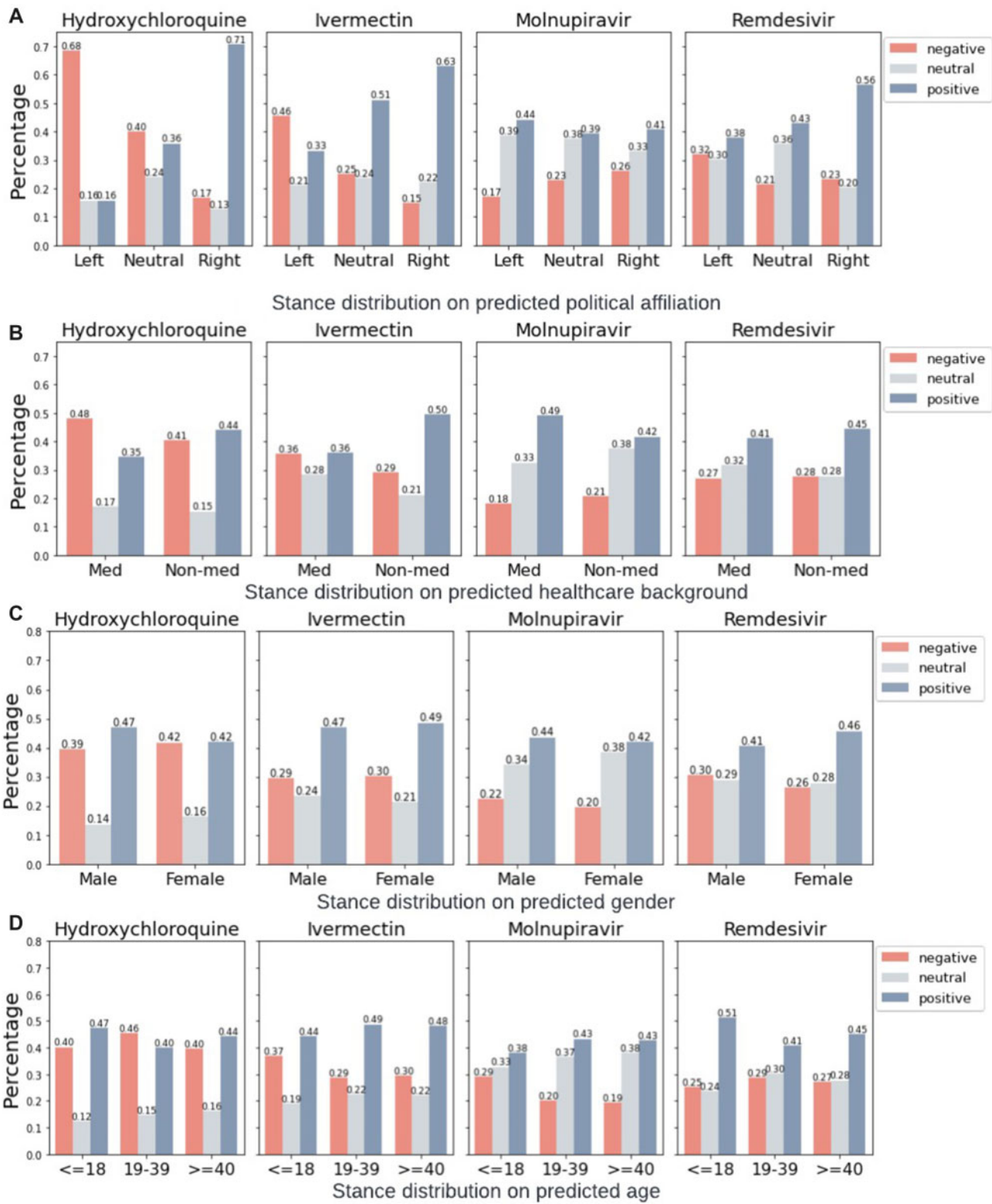

**Figure 6.** Stance distribution on predicted partisanship, age, and medical background for each drug. The exact numbers of tweets can be found in Supplementary Appendix SF.

Hydroxychloroquine because they didn't want the United States to be back to normal before the 2020 presidential election. Some other less frequent observations include that Hydroxychloroquine supporters were disappointed at the inconsistent stances of the US offi- cials (eg, Trump Administration vs. Centers for Disease Control and Prevention [CDC]), which deepened their distrust towards the government and made them believe that the government is hiding the treatment from them. Some Ivermectin supporters argued that peo-ple who could not get vaccinations because of immunodeficiency should be allowed to have other choices, that is, Ivermectin. Some Twitter users without a strong standpoint seemed to be irritated by the strong censorship of off-label drug use mentions on major media such as Facebook, YouTube, and Twitter. As a result, they may have perceived that authorities were over-reacting and attempting to politicize the drug.

Our study also shows that although Hydroxychloroquine has long been proven not to provide meaningful benefits for COVID-19 patients in September 2020 (the junction point of waves 1 and 2),[46] the proportion of its supporting tweets continued to grow. Some people believed that Hydroxychloroquine was evaluated unfairly and suspected that the Democrats used science for political purposes. This demonstrates that scientific reports on drugs will not end misinformation. As the number of prescriptions of Ivermectin for use by humans in the United States has increased to 24 times the prepandemic level, and prescriptions for veterinary use are also increasing,[47] we call for attention from public health agencies, policymakers, and health professionals to our findings to develop better public health strategies to reduce misinformation at an early stage. These strategies may include tailored education to individuals/groups with different demographic backgrounds and engaging people with different political views to depoliticize science and promote the spread of accurate information online.

Looking into the demographic traits of supporters and opponents of each drug, we found that the public opinion towards COVID-19-related drugs is highly divided in terms of political partisanship. This observation confirmed findings from previous work that partisanship affected public reaction to potential COVID-19 treatments.[48–51] While partisanship has long affected public responses to social issues and politicians, it is still striking to see how much this polarization has affected the public's search for potential cures during a health crisis. Previous studies have also shown that political, media, and technological forces have driven us into isolated and like-minded camps hostile to outside views.[52,53] We propose that public agencies, social platforms, and the government should take actions to engage people with different political opinions to value science and promote the spread of true information online for the common social good to reduce such effects of echo chambers.[54]

In addition to the findings and suggestions mentioned above, our open-source models and innovative pipeline can be used for similar social media-based public opinion analyses. In detail, we combined BERT-based stance detection and text clustering to extract fine-grained top rationales from the proponents and opponents of each drug. This approach allows researchers to summarize better and understand the public opinions toward drug use. In addition, we combined different existing modules to infer the demographic traits of Twitter users to study the relationship between these attributes and user stances. This pipeline allows researchers to understand the user background without human annotations. In practice, our methods characterize and capture complex user characteristics and social phenomena, which could provide important insights into developing better public health strategies to reduce misinformation at an early stage.

### Limitations

Public opinion analysis conducted on the Twitter platform may suffer from cohort bias, as the statistics on Statista show that Twitter does not cover a representative subset of the whole population.[55] Also, this study used geolocation information to identify US Twitter users, which is not always available and/or reliable. To mitigate this, we included 2 ways of geolocation extraction in our method, considering both geolocation sharing and self-reported location from user profiles (Supplementary Appendix SB). However, many Tweets from US users who didn't have access to a mobile phone with GPS or who preferred not to share their location information may have been missed from our analysis. While the ability to know where Twitter users are located is very valuable, the missingness may have altered the population distribution of the initial data and introduced potential bias. In addition, the M3 model has bias due to its training data, although the authors have applied debiasing techniques. This M3 model is also limited by encoding gender as a binary variable. The political partisanship inference method is also limited because it does not handle users who do not follow many politicians or users who like to follow politicians from both parties. Future work should develop more effective NLP solutions to facilitate opinion mining with fewer human annotations.

## CONCLUSION

This study developed a pipeline based on qualitative and quantitative analyses, leveraging modern NLP techniques such as BERT, as well as machine learning. Applying the pipeline on nearly 2 years of data, we filled in the research gap of understanding public perception towards COVID-19-related drugs. We found that the amount of public discourse on off-label drugs far exceeded that for non-off-label drugs during the pandemic. The reasons for supporting off-label drugs mostly came from hearsay, celebrity effects, and conspiracy theories. Political orientation and healthcare background are significantly associated with attitudes toward each drug. We call on public health agencies to develop targeted campaigns for safe medication use to avoid counterproductive effects. NLP-based tools can be leveraged to detect trends, monitor public opinions, and conduct demographic analyses on public discourse, effectively informing journalists, social media platforms, and government agencies of the public's thoughts.

## AUTHOR CONTRIBUTIONS

YH and HJ designed the study. LZ supervised the study design and implementation. JY provided critical feedback on the study design and implementation. YH retrieved and processed the data, curated the drug lexicon, conducted stance detection, trend analysis, geo analysis, and content analyses; HJ conducted the demographic analysis and helped with the content analysis. SL helped with literature review, stance detection evaluation, and content analysis. YH and HJ drafted the article. JY, JP, DB, and LZ provided critical instructions and reviews. All authors reviewed the article. YH takes responsibility for the integrity of the work.

## SUPPLEMENTARY MATERIAL

Supplementary material is available at *Journal of the American Medical Informatics Association Open* online.

## CONFLICT OF INTEREST STATEMENT

None declared.

## DATA AVAILABILITY

Data, source code, and pipeline tutorial of this paper are available at: https://github.com/ningkko/COVID-drug. The pretrained stance detection models can be downloaded at: https://huggingface.co/ningkko/drug-stance-bert.

## REFERENCES

1. Moshkovits I, Shepshelovich D. Emergency use authorizations of COVID-19-related medical products. *JAMA Intern Med* 2022; 182 (2): 228–9.
2. Zhai MZ, Lye CT, Kesselheim AS. Need for transparency and reliable evidence in emergency use authorizations for coronavirus disease 2019 (COVID-19) therapies. *JAMA Intern Med* 2020; 180 (9): 1145–6.
3. Young BE, Ong SWX, Kalimuddin S, et al.; Singapore 2019 Novel Coronavirus Outbreak Research Team. Epidemiologic features and clinical course of patients infected with SARS-CoV-2 in Singapore. *JAMA* 2020; 323 (15): 1488–94.
4. Wu Z, McGoogan JM. Characteristics of and important lessons from the coronavirus disease 2019 (Covid-19) outbreak in China: summary of a report of 72 314 cases from the Chinese Center for Disease Control and Prevention. *JAMA* 2020; 323 (13): 1239–42.
5. Pers YM, Padern G. Revisiting the cardiovascular risk of hydroxychloroquine in RA. *Nat Rev Rheumatol* 2020; 16 (12): 671–2.
6. Kalil AC. Treating COVID-19—off-label drug use, compassionate use, and randomized clinical trials during pandemics. *JAMA* 2020; 323 (19): 1897–8.
7. Ali S. Combatting against Covid-19 & misinformation: a systematic review. *Hu Arenas* 2022; 5 (2): 337–52.
8. Hamamsy T, Bonneau R. Twitter activity about treatments during the COVID-19 pandemic: case studies of remdesivir, hydroxychloroquine, and convalescent plasma. *medRxiv*. Published online July 14, 2020. doi: 10.1101/2020.06.18.20134668.
9. Marcon AR, Caulfield T. The Hydroxychloroquine Twitter War: a case study examining polarization in science communication. *First Monday* 2021; 26 (10). doi: 10.5210/fm.v26i10.11707.
10. Do T, Nguyen D, Le A, et al. Understanding public opinion on using hydroxychloroquine for COVID-19 treatment via social media. In: *15th International Conference on Health Informatics (HEALTHINF'22)*. 2021; Vienna, Sydney. doi: 10.5220/0010884200003123.
11. Grootendorst M. BERTopic: neural topic modeling with a class-based TF-IDF procedure. *ArXiv220305794 Cs*. Published online March 11, 2022. http://arxiv.org/abs/2203.05794. Accessed March 23, 2022.
12. Chew C, Eysenbach G. Pandemics in the age of Twitter: content analysis of tweets during the 2009 H1N1 outbreak. *PLoS One* 2010; 5 (11): e14118.
13. Oyeyemi SO, Gabarron E, Wynn R. Ebola, Twitter, and misinformation: a dangerous combination? *BMJ* 2014; 349: g6178.
14. Cinelli M, Quattrociocchi W, Galeazzi A, et al. The COVID-19 social media infodemic. *Sci Rep* 2020; 10 (1): 16598.
15. Cotfas LA, Delcea C, Gherai R, Roxin I. Unmasking people's opinions behind mask-wearing during COVID-19 pandemic—a Twitter stance analysis. *Symmetry* 2021; 13 (11): 1995.
16. Ebeling R, Sáenz CAC, Nobre J, Becker K. Analysis of the influence of political polarization in the vaccination stance: the Brazilian COVID-19 scenario. In: *Proceedings of the AAAI 16th International Conference on Web and Social Media (ICWSM'22)*. 2022; Atlanta, GA. doi: 10.48550/arXiv.2110.03382.
17. Lyu JC, Han EL, Luli GK. COVID-19 vaccine–related discussion on Twitter: topic modeling and sentiment analysis. *J Med Internet Res* 2021; 23 (6): e24435.
18. Muric G, Wu Y, Ferrara E. COVID-19 vaccine hesitancy on social media: building a public Twitter data set of antivaccine content, vaccine misinformation, and conspiracies. *JMIR Public Health Surveill* 2021; 7 (11): e30642.
19. Yeung N, Lai J, Luo J. Face off: polarized public opinions on personal face mask usage during the COVID-19 pandemic. In: *2020 IEEE International Conference on Big Data (Big Data)*. 2020: 4802–10.
20. Kruspe A, Häberle M, Kuhn I, Zhu XX. Cross-language sentiment analysis of European Twitter messages during the COVID-19 pandemic. In: *Proceedings of the 1st Workshop on NLP for COVID-19 at ACL*. 2020. https://aclanthology.org/2020.nlpcovid19-acl.14.
21. Klein AZ, Magge A, O'Connor K, Flores Amaro JI, Weissenbacher D, Gonzalez Hernandez G. Toward using Twitter for tracking COVID-19: a natural language processing pipeline and exploratory data set. *J Med Internet Res* 2021; 23 (1): e25314.
22. Crocamo C, Viviani M, Famiglini L, Bartoli F, Pasi G, Carrà G. Surveilling COVID-19 emotional contagion on Twitter by sentiment analysis. *Eur Psychiatry* 2021; 64 (1): e17.
23. Ebeling R, Sáenz C, Nobre C, Becker J. The effect of political polarization on social distance stances in the Brazilian COVID-19 scenario. *J Inform Data Manage* 2021; 12 (1). doi: 10.5753/jidm.2021.1889.
24. Duong V, Luo J, Pham P, Yang T, Wang Y. The ivory tower lost: how college students respond differently than the general public to the COVID-19 pandemic. In: *2020 IEEE/ACM International Conference on Advances in Social Networks Analysis and Mining (ASONAM)*. 2020: 126–30. doi: 10.1109/ASONAM49781.2020.9381379.
25. Dong E, Du H, Gardner L. An interactive web-based dashboard to track COVID-19 in real time. *Lancet Infect Dis* 2020; 20 (5): 533–4.
26. Lopez CE, Gallemore C. An augmented multilingual Twitter dataset for studying the COVID-19 infodemic. *Soc Netw Anal Min* 2021; 11 (1): 102.
27. Chen E, Lerman K, Ferrara E. Tracking social media discourse about the COVID-19 pandemic: development of a public Coronavirus Twitter data set. *JMIR Public Health Surveill* 2020; 6 (2): e19273.
28. Antiviral Therapy. COVID-19 Treatment Guidelines. https://www.covid19treatmentguidelines.nih.gov/therapies/antiviral-therapy/. Accessed March 30, 2022.
29. The Center for Drug Evaluation and Research | Coronavirus (COVID-19) | Drugs | FDA. Published July 2, 2021. https://www.fda.gov/drugs/emergency-preparedness-drugs/coronavirus-covid-19-drugs. Accessed March 30, 2022.
30. Loureiro D, Barbieri F, Neves L, Espinosa Anke L, Camacho-Collados J. TimeLMs: diachronic language models from Twitter. In: *Proceedings of the 60th Annual Meeting of the Association for Computational Linguistics: System Demonstrations*. 2022: 251–60; Dublin, Ireland. https://aclanthology.org/2022.acl-demo.25. Accessed June 2, 2022.
31. Mutlu EC, Oghaz T, Jasser J, et al. A stance data set on polarized conversations on Twitter about the efficacy of hydroxychloroquine as a treatment for COVID-19. *Data Brief* 2020; 33: 106401.
32. Jiang H, Hua Y, Beeferman D, Roy D. Annotating the Tweebank corpus on named entity recognition and building NLP models for social media analysis. In: *Proceedings of the 13th Edition of the Language Resources and Evaluation Conference (LREC'22)*. 2022; Marseille, France. doi: 10.48550/arXiv.2201.07281.
33. Qi P, Zhang Y, Zhang Y, et al. Stanza: a python natural language processing toolkit for many human languages. In: *Proceedings of the 58th Annual Meeting of the Association for Computational Linguistics: System Demonstrations*. 2020: 101–8.
34. Demszky D, Garg N, Voigt R, et al. Analyzing polarization in social media: method and application to Tweets on 21 mass shootings. In: *Proceedings of the 2019 Conference of the North American Chapter of the Association for Computational Linguistics: Human Language Technologies, Volume 1 (Long and Short Papers)*. 2019: 2970–3005; Minneapolis. doi: 10.18653/v1/N19-1304.
35. Li M, Hua Y, Liao Y, et al. Tracking the impact of COVID-19 and lockdown policy on public mental health using social media. *Research Square*. Published online March 30, 2022. doi: 10.21203/rs.3.rs-1498436/v1.
36. Wang Z, Hale S, Adelani DI, et al. Demographic inference and representative population estimates from multilingual social media data. In: *The World Wide Web Conference. WWW '19*. Association for Computing Machinery. 2019: 2056–67; San Francisco.
37. Niburski K, Niburski O. Impact of Trump's promotion of unproven COVID-19 treatments and subsequent internet trends: observational study. *J Med Internet Res* 2020; 22 (11): e20044.
38. DGHS drops Ivermectin, Doxycycline from Covid-19 treatment; ICMR rules unchanged—Coronavirus Outbreak News. https://www.indiatoday.in/coronavirus-outbreak/story/revised-health-ministry-guidelines-stop-usage-of-ivermectin-doxycycline-in-covid-treatment-1811809-2021-06-07. Accessed March 30, 2022.
39. Fact Check-Japan has not authorized ivermectin to treat COVID-19 or revoked a vaccine mandate. *Reuters*. Published January 17, 2022. https://

www.reuters.com/article/factcheck-japan-ivermectinandmandate-idUSL1N2TX1GK. Accessed March 30, 2022.
40. Fact Check-Australian hospitals are not treating vaccinated COVID-19 patients with ivermectin. *Reuters*. Published September 22, 2021. https://www.reuters.com/article/factcheck-australia-ivermectin-idUSL1N2-QO0P1. Accessed March 30, 2022.
41. Health AGD of. COVID-19 treatments. Australian Government Department of Health. Published February 14, 2021. https://www.health.gov.au/health-alerts/covid-19/treatments. Accessed March 30, 2022.
42. Global Twitter user distribution by gender 2022. Statista. https://www.statista.com/statistics/828092/distribution-of-users-on-twitter-worldwide-gender/. Accessed April 20, 2022.
43. U.S. Twitter reach by age group 2021. Statista. https://www.statista.com/statistics/265647/share-of-us-internet-users-who-use-twitter-by-age-group/. Accessed March 30, 2022.
44. Kamiński M, Szymańska C, Nowak JK. Whose Tweets on COVID-19 gain the most attention: celebrities, political, or scientific authorities? *Cyberpsychol Behav Soc Netw* 2021; 24 (2): 123–8.
45. Vosoughi S, Roy D, Aral S. The spread of true and false news online. *Science* 2018; 359 (6380): 1146–51.
46. Hydroxychloroquine does not benefit adults hospitalized with COVID-19. National Institutes of Health (NIH). Published November 9, 2020. https://www.nih.gov/news-events/news-releases/hydroxychloroquine-does-not-benefit-adults-hospitalized-covid-19. Accessed April 5, 2022.
47. HAN Archive—00449 | Health Alert Network (HAN). Published September 21, 2021. https://emergency.cdc.gov/han/2021/han00449.asp. Accessed April 5, 2022.
48. Sanchez C, Dunning D. The anti-Scientists bias: The role of feelings about scientists in COVID-19 attitudes and behaviors. *J Appl Soc Psychol* 2021; 51 (4): 461–73.
49. Brunell TL, Maxwell SP. How partisanship affected public reaction to potential treatments for COVID-19. *World Med Health Policy* 2020; 12 (4): 482–6.
50. Havey N. Partisan public health: how does political ideology influence support for COVID-19 related misinformation? | SpringerLink. Published 2020. https://link.springer.com/article/10.1007/s42001-020-00089-2. Accessed April 6, 2022.
51. Barnett ML, Gaye M, Jena AB, Mehrotra A. Association of county-level prescriptions for hydroxychloroquine and ivermectin with county-level political voting patterns in the 2020 US Presidential Election. *JAMA Intern Med* 2022; 182 (4): 452–4.
52. Sunstein C. The Law of Group Polarization. *Law & Economics of Working Paper*. Published online December 1, 1999. https://chicagounbound.uchicago.edu/law_and_economics/542.
53. Gillani N, Yuan A, Saveski M, Vosoughi S, Roy D. Me, My Echo Chamber, and I: introspection on social media polarization. In: *Proceedings of the 2018 World Wide Web Conference*. WWW '18. International World Wide Web Conferences Steering Committee; 2018: 823–31; Lyon, France.
54. Saveski M, Beeferman D, McClure D, Roy D. Engaging Politically Diverse Audiences on Social Media. In: *Proceedings of the AAAI 16th International Conference on Web and Social Media (ICWSM'22)*. 2022; Atlanta, GA. doi: 10.48550/arXiv.2111.02646.
55. Mislove A, Lehmann S, Ahn YY, Onnela JP, Rosenquist J. Understanding the demographics of Twitter Users. In: Proceedings of the Fifth International AAAI Conference on Web and Social Media (ICWSM-11); 2011; Vol. 5: 554–7; Barcelona, Catalonia.